\documentclass[number,preprint,review,12pt,3p]{elsarticle}

\usepackage{amssymb}

\usepackage{url}
\usepackage{lineno}

\biboptions{sort&compress}

\journal{Nucl. Instrum. Meth. A}

\begin{document}

\begin{frontmatter}



\title{Demonstration of a 25-picosecond single-photon time resolution with gaseous photomultiplication}


\author[KEK,SOKENDAI,Nagoya]{K.~Matsuoka\corref{cor1}}
\ead{matsuoka@post.kek.jp}
\cortext[cor1]{Corresponding author.}
\author[Nagoya]{R.~Okubo}
\author[Nagoya]{Y.~Adachi}

\affiliation[KEK]{organization={High Energy Accelerator Research Organization (KEK)},
            city={Tsukuba},
            postcode={305-0801}, 
            country={Japan}}
\affiliation[SOKENDAI]{organization={The Graduate University for Advanced Studies (SOKENDAI)},
            city={Hayama},
            postcode={240-0193},
            country={Japan}}
\affiliation[Nagoya]{organization={Graduate School of Science, Nagoya University},
            city={Nagoya},
            postcode={464-8602}, 
            country={Japan}}

\begin{abstract}

Photosensitive gaseous detectors with a simple photoelectron multiplication mechanism of resistive plate chambers (RPCs) are expected to have both a large photocoverage and an excellent time resolution and to be low-cost.
To demonstrate the time resolution of the RPC-based photodetectors, we built a prototype detector with a LaB$_6$ photocathode.
It was tested with a picosecond pulse laser and the intrinsic time resolution for single photons was measured to be $25.0 \pm 1.1$~ps at a gain of $3.3\times 10^6$.

\end{abstract}




\end{frontmatter}


\section{Introduction}
Scale-up of the size and performance of detectors is inevitable for future nuclear and particle experiments to exploit the physics frontier.
Regarding particle detectors which detect Cherenkov or scintillation light, photodetectors with a high photodetection efficiency, a large photocoverage, and an excellent time resolution are commonly required~\cite{photodetectors}.
The cost, which increases with the size and performance, is also a serious concern.

Keeping the time resolution while enlarging the photocoverage is challenging mainly due to increase of the device capacitance, which increases the noise and hence time jitter, for solid-state photodetectors and due to increase of the transit time spread (TTS) for vacuum-based photomultiplier tubes (PMTs).
Among silicon photodetectors, the best single-photon time resolution to date is 7.8~ps full width at half maximum (FWHM)\footnote{In this paper only single-photon time resolution is discussed and it is expressed in the standard deviation of Gaussian distribution unless otherwise specified. It should be noted that in most cases the measured time distributions of output signals have a non-negligible tail component or even a second peak, which is not reflected in the quoted time resolutions.}, but it was achieved with a very small single-photon avalanche diode of 20~{\textmu}m diameter and a quenching circuit implemented in CMOS 65~nm technology~\cite{SPAD-QC}.
Among PMTs, micro-channel-plate (MCP) PMTs have the best time resolution of about 30~ps with the channel diameter of 10~{\textmu}m~\cite{MCP-PMT_Belle2, MCP-PMT_Photonis, MCP-PMT_Photek}.
The planar geometry of MCP-PMTs enables to enlarge the photocoverage without deteriorating the TTS.
The largest MCP-PMT~\cite{LAPPD} has an effective area of 350~cm$^2$, while the fill factor is similar to or even smaller than other MCP-PMTs of a smaller shape because of the spacer sustaining the vacuum enclosure.
The high cost of MCP-PMTs can easily become a bottleneck to cover a large area with a number of MCP-PMTs.
Superconducting nanowire single-photon detectors (SNSPDs)~\cite{SNSPD} can outperform any other free-running single-photon detectors in terms of the time resolution.
The observed best time resolution is 2.6~ps FWHM~\cite{SNSPD_2.6ps}.
However, the scalability is a key challenge.
The largest SNSPD array to date has an active area of $0.96\times 0.96$~mm~\cite{SNSPD_kilopixel}.

Photosensitive gaseous detectors have advantages over the other types of photodetectors mainly in terms of the size and cost.
The sensitivity is, however, limited mostly to ultraviolet (UV) photons as one can only use photocathodes that are not vulnerable to the gas or some photosensitive gas mixtures~\cite{photosensitive_gas}.
Nevertheless, the UV sensitivity is suitable for Cherenkov and UV-scintillation detectors, and photosensitive gaseous detectors with cesium iodide (CsI) photocathodes have been widely used~\cite{PGD}.
With a planar uniform geometry and atmospheric pressure of the gas, they can be easily enlarged without diminishing the time resolution.
A time resolution of 44~ps for single photons was demonstrated by a PICOSEC-Micromegas detector~\cite{PICOSEC-Micromegas}.
As is the case for this detector, micro-pattern gaseous detectors (MPGDs) are used for the photoelectron multiplication in order to suppress photon and ion feedback to the photocathode.
Utilizing parallel-plate avalanche chambers~\cite{PPAC_CsI} and resistive plate chambers (RPCs)~\cite{RPC_CsI_1, RPC_CsI_2, RPC_CsI_3} were also investigated.
Though they could not be operated at a high gain without breakdown due to the photon and ion feedback problem~\cite{PGD}, they potentially outperform MPGDs in terms of the photodetection efficiency and time resolution.
That is because of the higher electric field near the photocathode, which enhances the quantum efficiency~\cite{QE_field}, enables prompt multiplication, leading to a lower probability of photoelectron attachment, and reduces the electron diffusion.
In addition, their simpler structures than MPGDs could help reduce the cost.

We aim to design an RPC-based photosensitive gaseous detector, which we call gaseous photomultiplier (GasPM), capable of having both a large photocoverage and an excellent time resolution in a lower-cost way than any other photodetectors.
As a first step, we made a prototype detector to demonstrate the capability of a good time resolution for single photons.
Test results for the prototype are shown in this paper.

\section{Design of GasPM}
\begin{figure}[tb]
	\centering 
	\includegraphics[width=0.5\textwidth]{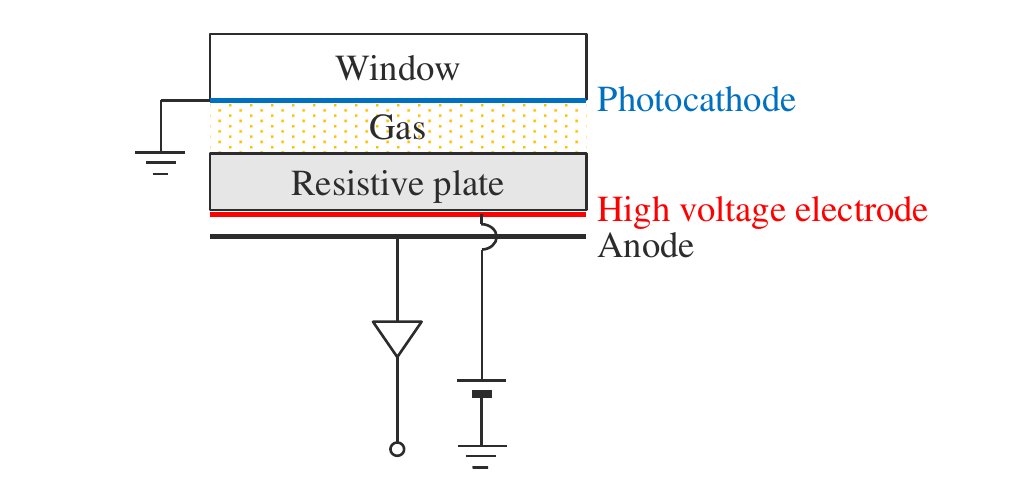}
	\caption{Schematic design of GasPM.}
	\label{fig:design_schematic}
\end{figure}

The schematic design of GasPM is shown in Fig.~\ref{fig:design_schematic}.
GasPM is composed of a planar photocathode on the inner side of a window for light input, a resistive plate parallel to the photocathode, a high voltage electrode with low resistivity, and a conductive anode pad.
A uniform electric field is generated in between the high voltage electrode and the photocathode, which is connected to ground.
The resistive plate has a relatively high electric resistivity and is positively charged up to have the same electric potential as the applied high voltage.
A suitable gas mixture is filled in the narrow gap of $\mathcal{O}$(100~{\textmu}m) between the window and resistive plate.
Photoelectrons from the photocathode are multiplied by the avalanche process in the gas, and the mirror current of the avalanche signal is induced on the anode pad.
The resistive plate works to suppress the growth of the avalanche into discharge as the electric potential on the local area of the resistive plate drops drastically.
The working principle for the electron multiplication is the same as RPCs~\cite{1stRPC}.

Focusing only on the time resolution, the first prototype of GasPM was designed to possess the following features:
\begin{itemize}
	\item Easy to assemble on a table by adopting a lanthanum hexaboride (LaB$_6$) photocathode, which can be handled in air.
	\item Low in risk of electric breakdown or discharge in the narrow gap by adopting high-resistive TEMPAX Float\textsuperscript{\textregistered} glass for the resistive plate.
\end{itemize}

LaB$_6$ has been known as a low work-function photoemitter~\cite{LaB6}.
The work-function of LaB$_6$ clean surface is 2.3--3.3~eV depending on the surface structure~\cite{LaB6_surface}.
When the surface is exposed to oxygen, forming one monolayer of oxide, the work-function increases by up to 1.4~eV ~\cite{LaB6_oxidized}.
For a thin film of LaB$_6$ after exposure to air, the quantum efficiency of the order of $10^{-5}$ for 4.7~eV photons was reported in Ref.~\cite{LaB6_QE}.
Though LaB$_6$ photocathodes are expected to have an extremely low quantum efficiency, they still work for our demonstration with single photons by illumination of UV laser.

The volume resistivity of TEMPAX Float glass is $10^{15}$~{\textohm}cm.
A drawback of this high resistivity is a long time constant to charge up the resistive plate, which diminishes the capability of high rate detection, but it is not an issue for our demonstration.

\begin{figure*}[tb]
	\centering 
	\includegraphics[width=\textwidth]{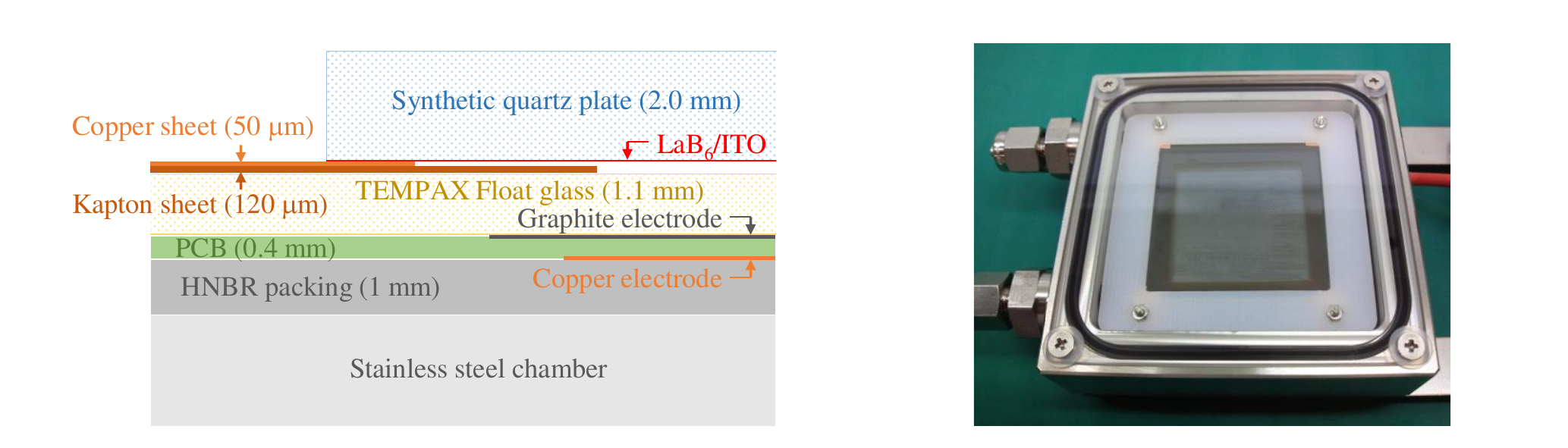}
	\caption{(Left) Cross-sectional schematic design of the GasPM prototype inside the stainless steel chamber.
		Only a part around the edge of the photocathode is shown in this figure.
		The thickness of each component is shown in parentheses.
		(Right) Picture of the GasPM prototype viewed from above the TEMPAX Float glass window.}
	\label{fig:design_prototype}
\end{figure*}

The design of the prototype is shown in Fig.~\ref{fig:design_prototype}.
Several layers are simply stacked inside a gas tight chamber made of stainless steel: from the bottom side, where the signal and high voltage feed-through terminals are located, a hydrogenated nitrile rubber (HNBR) packing to seal the feed-through terminals, a printed-circuit board (PCB) with a single copper electrode for the signal anode on the lower side and a graphite electrode with $\mathcal{O}$(1~k{\textohm}/sq.) sheet resistivity for the high voltage on the upper side, a 1.1-mm thick TEMPAX Float glass, a thin Kapton\textsuperscript{\textregistered} sheet for electric insulation, a thin copper sheet to contact the photocathode with the ground, and a synthetic quartz plate to place the photocathode.
The Kapton and copper sheets are also used as a spacer to define the gap between the photocathode and TEMPAX Float glass, which is 170-{\textmu}m thick and is filled with gas for electron multiplication.
The GasPM chamber is closed with an extra 2-mm thick TEMPAX Float glass window on top of the synthetic quartz window.
The extra window is used only for this prototype to make gas seal easy.
The synthetic quartz and TEMPAX Float glass windows are suitable to transmit UV laser to the photocathode; The transmittance is nearly constant at about 90\% down to the wavelength of 180~nm and 350~nm, respectively.
The sizes of the signal electrode, high voltage electrode, and photocathode are $31.2\times 31.2$~mm$^2$, $34.0\times 34.0$~mm$^2$, and $36.0\times 36.0$~mm$^2$, respectively, and the active area is limited to $30\times 30$~mm$^2$ by the Kapton sheet.
The GasPM prototype has only one output channel that covers the whole active area.

The LaB$_6$ film of the photocathode is underlaid with an indium tin oxide (ITO) film to assist in transmitting photocurrent and generating a uniform electric field because LaB$_6$ has a low electric conductivity.
With radio frequency (RF) magnetron sputtering, the ITO film was firstly deposited on one side of the synthetic quartz plate, and then the LaB$_6$ film was deposited on the ITO film with the four edges masked in order to bring the ITO film into contact with the copper sheet.
The target thickness of each film was determined to be 40~nm such that the absorptance of 400~nm light is maximized at the LaB$_6$ surface in contact with the gas.
It was calculated by using the exact expression of the absorptance~\cite{absorptance} and the refractive indices of LaB$_6$~\cite{index_LaB6} and ITO~\cite{index_ITO}.
The sputtering parameters for LaB$_6$ were determined as follows by reference to previous studies with direct current magnetron sputtering~\cite{LaB6_sputtering_power_flow, LaB6_sputtering_heat}: 100~W RF power, 2.0~sccm flow rate and 1.6~Pa pressure of the sputtering argon gas, and 400$^\circ$C substrate heating.
The ITO target was made of In$_2$O$_3$ (90~wt\%) and SnO$_2$ (10~wt\%).
Right after the fabrication of the LaB$_6$ photocathode, it was sensitive to 405~nm laser.
However, it became oxidized with time and turned out to be insensitive to photons above 400~nm wavelength.

The chamber is filled with a mixture of tetrafluoroethane (C$_2$H$_2$F$_4$, commercially available R134a) and sulfur hexafluoride (SF$_6$) gases, which have been widely used for RPCs.
GasPM is operated in avalanche mode, and SF$_6$ is essential to reduce streamer contamination~\cite{RPC_SF6}.
R134a and SF$_6$ were flowed at 2.25~cc/min and 0.25~cc/min, respectively, by mass flow controllers.
Fresh gas at the atmospheric pressure and the room temperature was always fed into the GasPM chamber .

With this prototype, it was confirmed that the high voltage can be applied up to 4.0~kV without any problem of electric breakdown or discharge.
Throughout measurements for the performance evaluation described hereinafter, 3.0~kV was applied, corresponding to 176~kV/cm.

\section{Expected performance of the GasPM prototype}
\begin{figure*}[tb]
	\begin{minipage}{0.49\textwidth}
		\centering 
		\includegraphics[width=\textwidth]{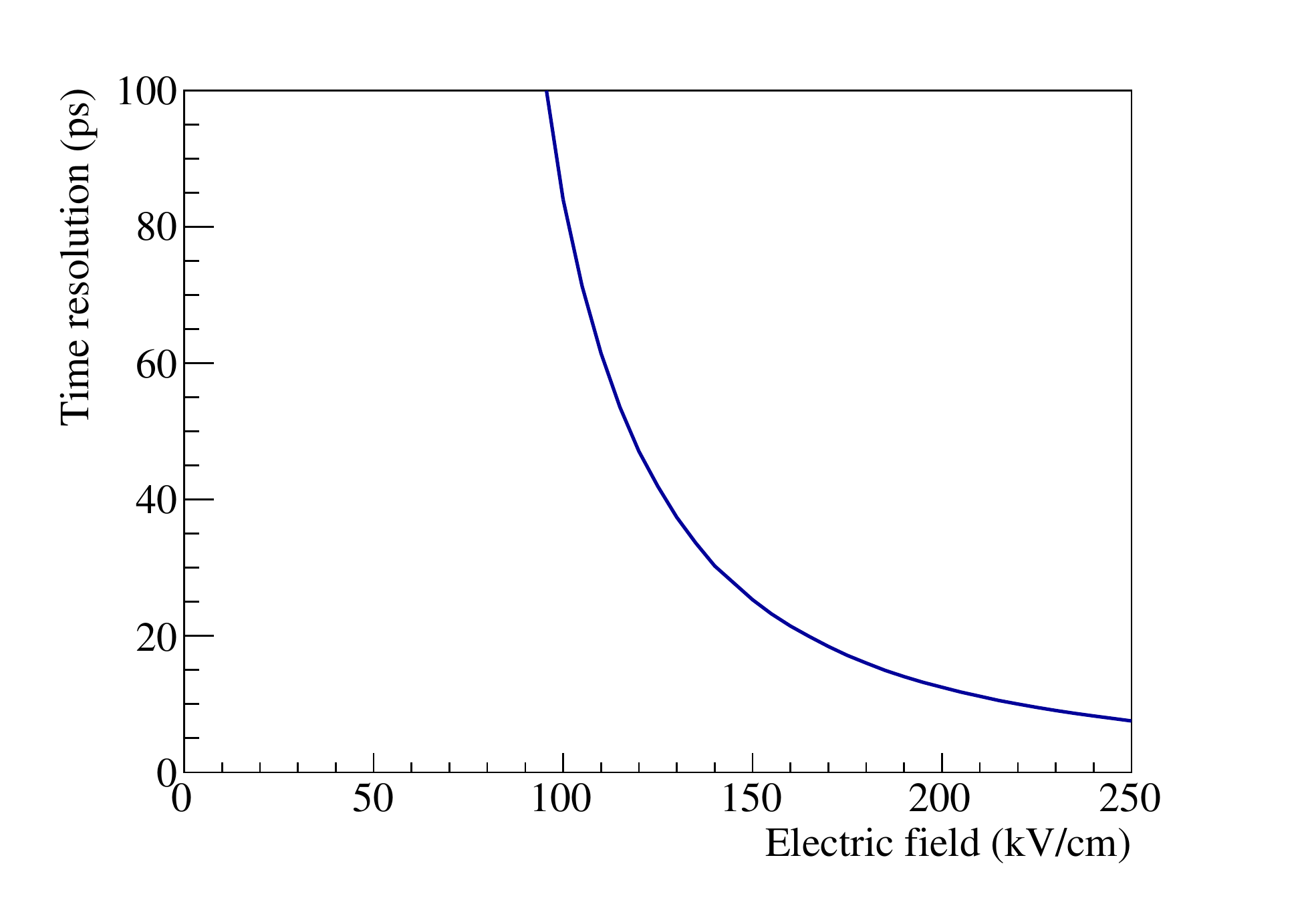}
	\end{minipage}
	\hspace{0.02\textwidth}
	\begin{minipage}{0.49\textwidth}
		\centering 
		\includegraphics[width=\textwidth]{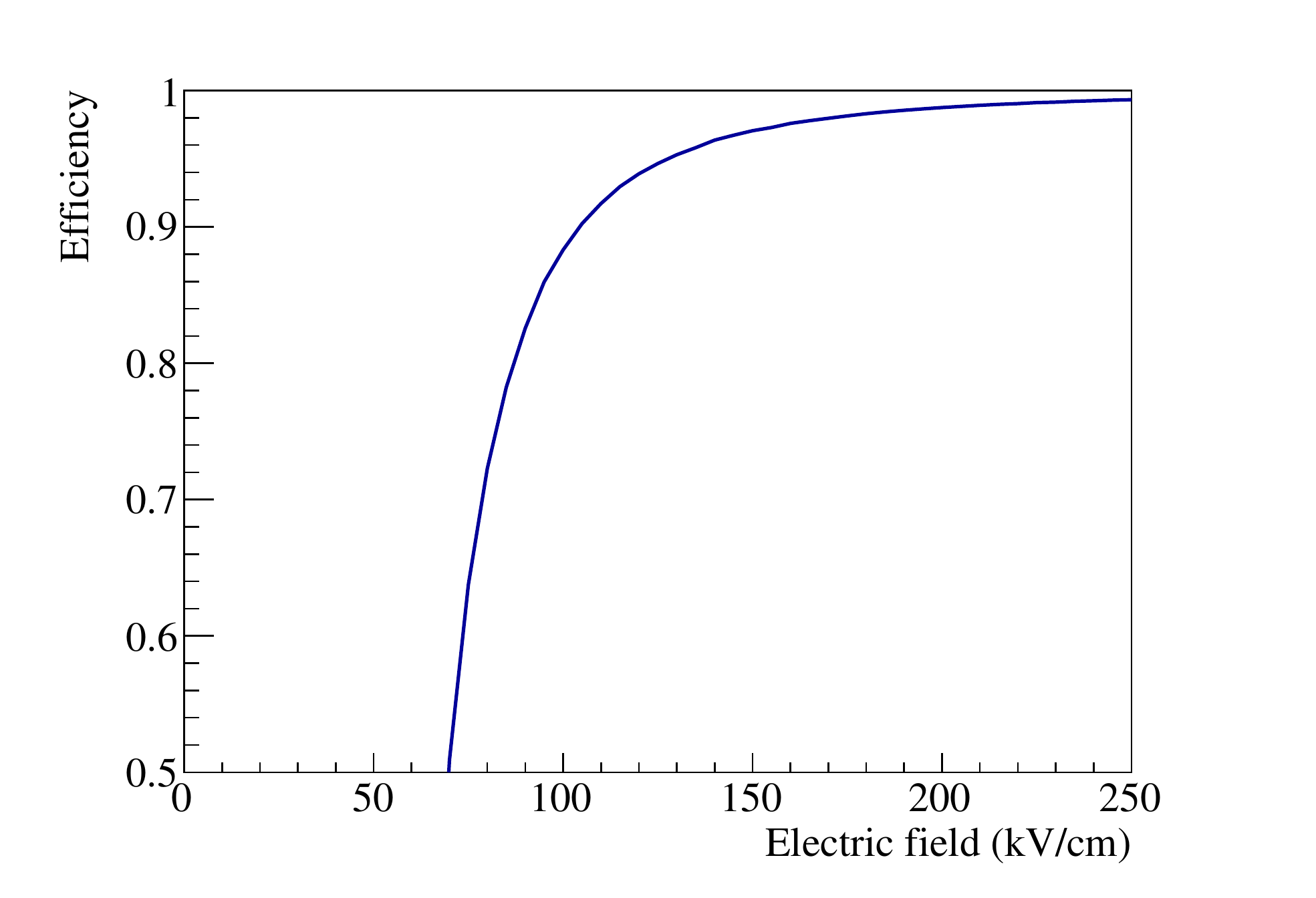}
	\end{minipage}
	\caption{Estimated intrinsic time resolution (left) and collection efficiency (right) as a function of the applied electric field for the GasPM prototype.}
	\label{fig/estimation}
\end{figure*}

GasPM is expected to have a good time resolution owing to the fast avalanche process in the narrow gap.
The intrinsic time resolution ($\sigma_t$) for single photons is roughly expressed as~\cite{RPC_simulation}
\begin{equation} \label{eq:resolution}
	\sigma_t = \frac{1.28}{(\alpha - \eta) v},
\end{equation}
where $\alpha$ and $\eta$ are the first Townsend and attachment coefficients, respectively, and $v$ is the electron drift velocity for the gas mixture.
A high electric field is demanded to reduce $\sigma_t$ as $\alpha - \eta$ and $v$ increase with the electric field.
The collection efficiency ($\varepsilon$), that is, the probability that a photoelectron from the photocathode is not attached by the gas and leads to an avalanche signal,
\begin{equation} \label{eq:efficiency}
	\varepsilon = 1 - \frac{\eta}{\alpha},
\end{equation}
is also expected to be high at a high electric field.
However, the maximum electric field will be limited by the problem of the photon and ion feedback.

The estimated $\sigma_t$ and $\varepsilon$ for the GasPM prototype from Eqs.~(\ref{eq:resolution}) and (\ref{eq:efficiency}) are shown in Fig.~\ref{fig/estimation}.
The parameters, $\alpha$, $\eta$, and $v$, for the gas mixture at 1~atm and 20$^\circ$C are computed with a Monte Carlo simulation program, MAGBOLTZ~2~\cite{Magboltz2}.
At 176~kV/cm, $\sigma_t = 17$~ps and $\varepsilon = 98$\%.
These are also confirmed by a Monte Carlo simulation of the electron multiplication process using Garfield++~\cite{Garfield++}.

GasPM also works as an RPC and is sensitive to charged particles passing through the gas gap.
The efficiency of the GasPM prototype detecting a minimum ionizing particle is estimated using the expression in Ref.~\cite{RPC_efficiency} to be about 0.6.
Therefore, cosmic muons cause random noise for the performance evaluation of the GasPM prototype.
The cosmic noise rate is estimated to be about 0.1~Hz.
Another type of random noise could be spurious pulses, which are practically well-known to occur in all gaseous detectors but are not well understood.
The spurious pulse rate has to be checked with the GasPM prototype.
The dark noise rate due to thermionic electron emission from the LaB$_6$ photocathode should be negligibly low considering the quite low quantum efficiency and insensitivity to photons above 400~nm wavelength.

\section{Performance of the GasPM prototype}
\subsection{Measurement setup}
The performance of the GasPM prototype for single photons was evaluated using a picosecond pulse diode laser with a center wavelength of 375~nm.
The time structure of the laser pulse was measured by the manufacturer of the laser module using a digital sampling optical oscilloscope (streak camera) with 11~ps FWHM impulse response.
It was close to a Gaussian with $22.3 \pm 0.5$~ps standard deviation.
Therefore, the pulse width of the laser excluding the oscilloscope response is estimated to be $21.8 \pm 0.5$~ps.
The GasPM prototype was placed inside a dark box, and the laser was spotted through an optical fiber on a $6\times 6$~mm$^2$ area around the center of the photocathode delimited by a light-blocking mask.
The laser pulse repetition rate was 100~MHz, and the average power of the laser at the fiber edge was about 1~mW.

The GasPM signal and the synchronous output signal from the laser module were recorded by a digitizer (DRS4~\cite{DRS4} evaluation board), which has a sampling rate of 5~Gsamples/sec and an analog bandwidth of 700~MHz.
The input channels of the digitizer are terminated with 50~\textohm.
The GasPM signal was amplified with a 37~dB wideband (0.5--1500~MHz) low-noise amplifier followed by a 10~dB attenuator, and hence 27~dB in total.
The data were recorded when the GasPM signal pulse height exceeded a $-15$~mV threshold.
The time resolution of the readout system was evaluated by a test pulse to be $14.0\pm0.3$~ps.

\subsection{Detector response to single photons}
The detection rate of the GasPM signals in coincidence with the laser synchronous output signal was 0.02~Hz.
The rate was quite low, reflecting an extremely low quantum efficiency of the LaB$_6$ photocathode, and thus all the signals are regarded as originating from single photoelectrons.
Random noise was also observed, and the rate fluctuated within a range of 0.3--1~Hz.
Therefore, the random noise is attributed mainly to spurious pulses.

\begin{figure}[tb]
	\centering
	\includegraphics[width=0.45\textwidth]{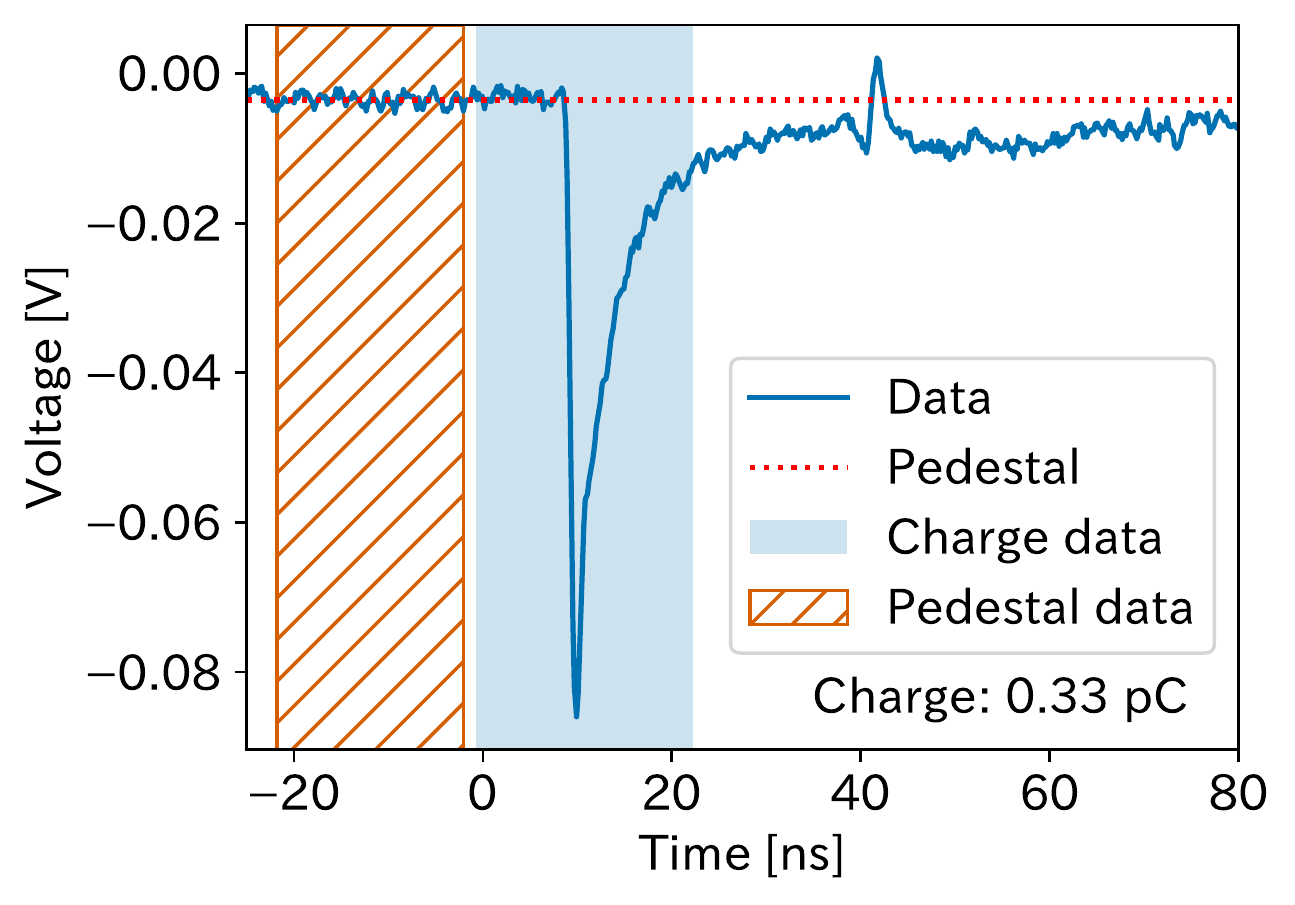}
	\caption{Typical single-photon signal of the GasPM prototype recorded by the digitizer.
		It was amplified by 27~dB.
		The red dotted line represents the pedestal, which is calculated from the sampling points in the red-orange hatched region.
		The net charge of this signal at the GasPM output without amplification is 0.33~pC, which is calculated from the sampling points in the blue filled region.
		The small pulse of the opposite polarity around 40~ns is a reflection of the first pulse.}
	\label{fig:waveform_signal}
\end{figure}
\begin{figure}[tb]
	\centering
	\includegraphics[width=0.45\textwidth]{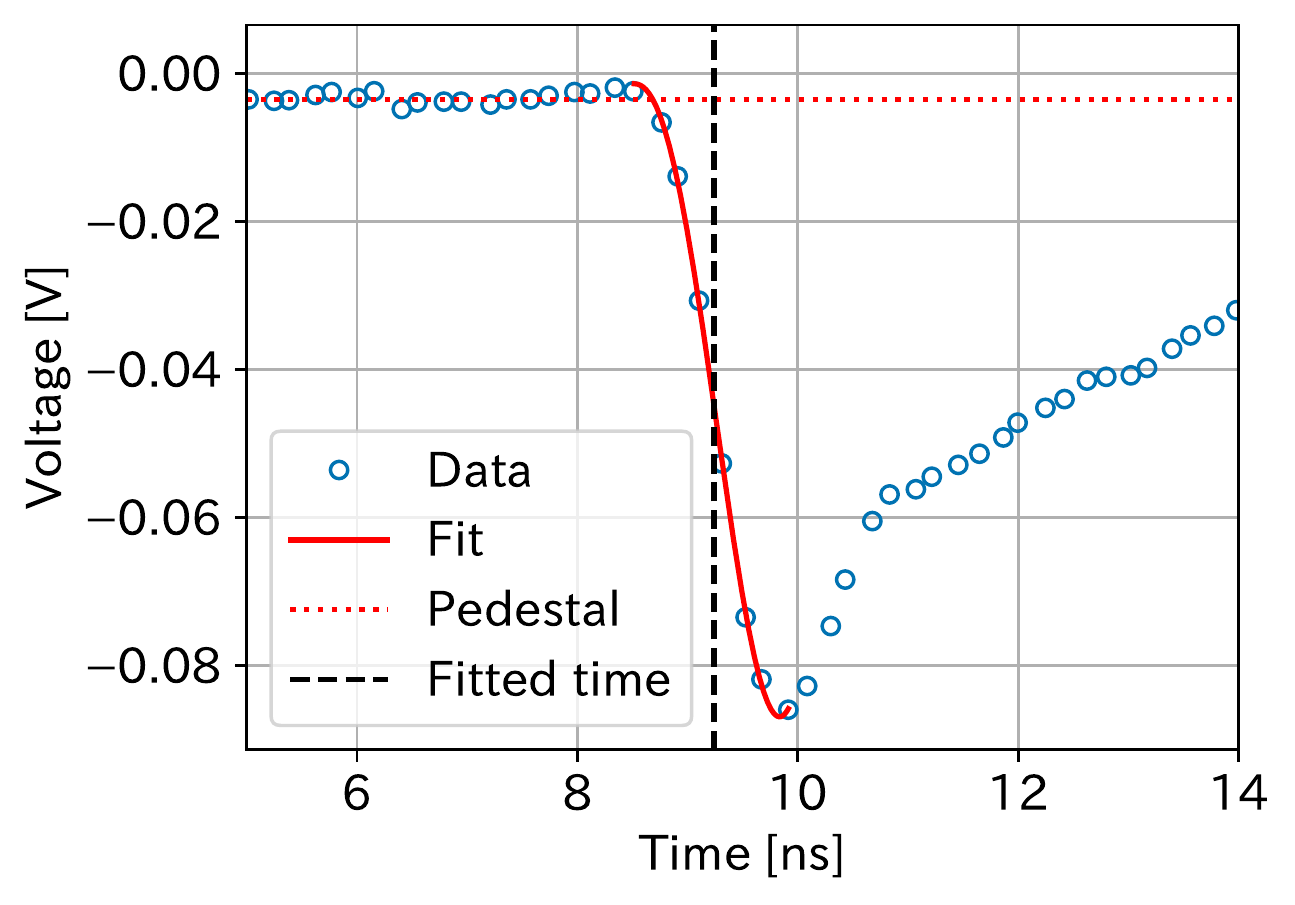}
	\caption{Rising edge of a single-photon signal of the GasPM prototype recorded by the digitizer.
		On top of the blue sampling points, the red dotted line for the pedestal and the red solid line for the fitted sixth-order polynomial function are superimposed.
		The black dashed line represents $t_\mathrm{sig}$.}
	\label{fig:waveform_fitting}
\end{figure}

A typical signal on the laser timing is shown in Fig.~\ref{fig:waveform_signal}.
It steeply rises in 1~ns and decays with a time constant of 4.3~ns, which is dictated mostly by the capacitance between the GasPM electrodes.
After the fast pulse from the electron drift, a much longer tail component remains, which is attributed to the ion drift.
The pedestal is calculated for each signal using the sampling points before the fast pulse.
The output charge or gain is calculated only for the electron component by integrating the sampling points from $t_\mathrm{sig} - 10$~ns to $t_\mathrm{sig} + 13$~ns, where $t_\mathrm{sig}$ is the time when the rising edge of the pulse reaches the half of the maximum voltage.
To deduce $t_\mathrm{sig}$, the rising edge is fitted by a sixth-order polynomial function as shown in Fig.~\ref{fig:waveform_fitting}.
The time origin is defined by the time of the laser synchronous output signal, which is deduced in the same way as $t_\mathrm{sig}$.

\begin{figure*}[tb]
	\begin{minipage}{0.49\textwidth}
		\centering
		\includegraphics[width=0.95\textwidth]{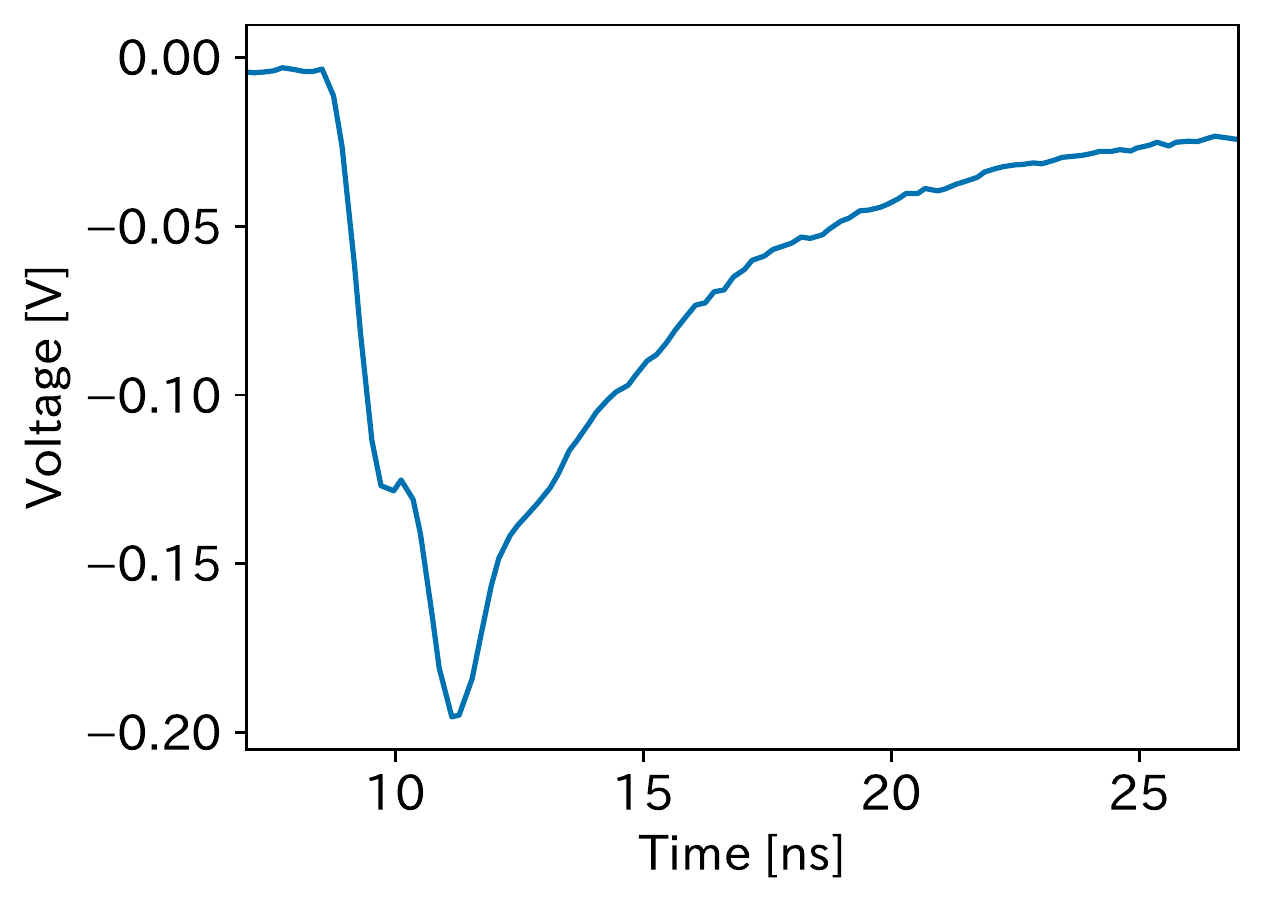}
	\end{minipage}
	\begin{minipage}{0.49\textwidth}
		\centering
		\includegraphics[width=0.95\textwidth]{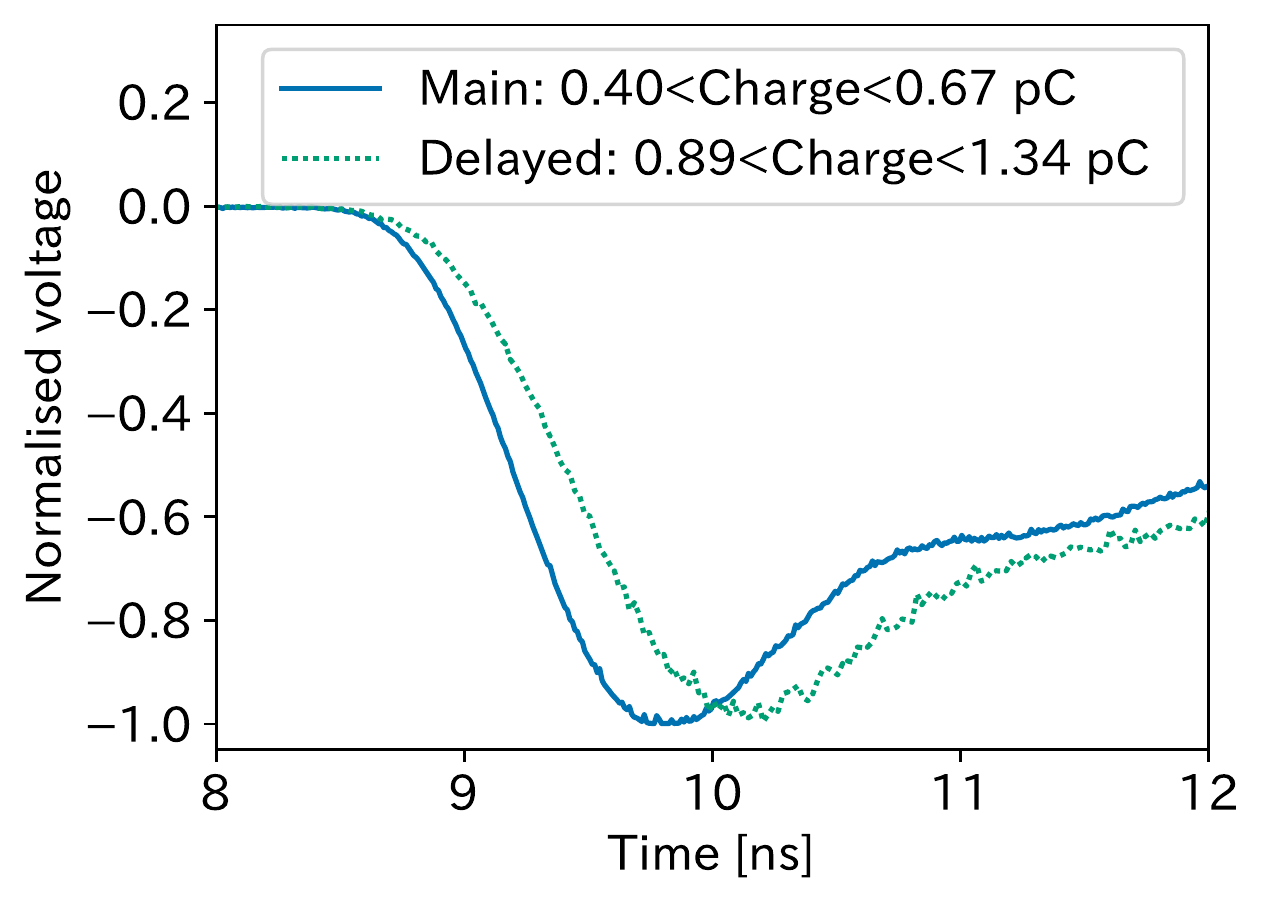}
	\end{minipage}
	\caption{(Left) Example of a double-peaked pulse.
		(Right) Averaged shapes of the signal pulses in the two output charge ranges.
		Each pulse height is normalized in this plot.}
	\label{fig:waveform_delayed}
\end{figure*}

Among the signals, some double-peaked pulses as shown in Fig.~\ref{fig:waveform_delayed} (left) were found.
We consider that the double peak is composed of two overlapping successive pulses: the first one from an ordinary avalanche and the second one from another avalanche due to photon feedback.
If the second pulse occurs promptly after the first pulse, the two pulses cannot be resolved but are observed as a single pulse of a larger height.
Moreover, the risetime of the two overlapping pulses with such a subtle temporal difference becomes slower.
Actually two distinct pulse heights and risetimes were observed for the signals on the laser timing.
Figure~\ref{fig:waveform_delayed} (right) shows the difference of the risetimes for the different output charge.
The larger pulse has the slower risetime, and the measured $t_\mathrm{sig}$ is slightly delayed by about 0.2~ns compared to the smaller pulse.
The signals with the smaller and steeper pulse shape are called main laser signal and those with the larger and slower pulse shape are called delayed laser signal hereafter.

\begin{figure}[tb]
	\centering
	\includegraphics[width=0.5\textwidth]{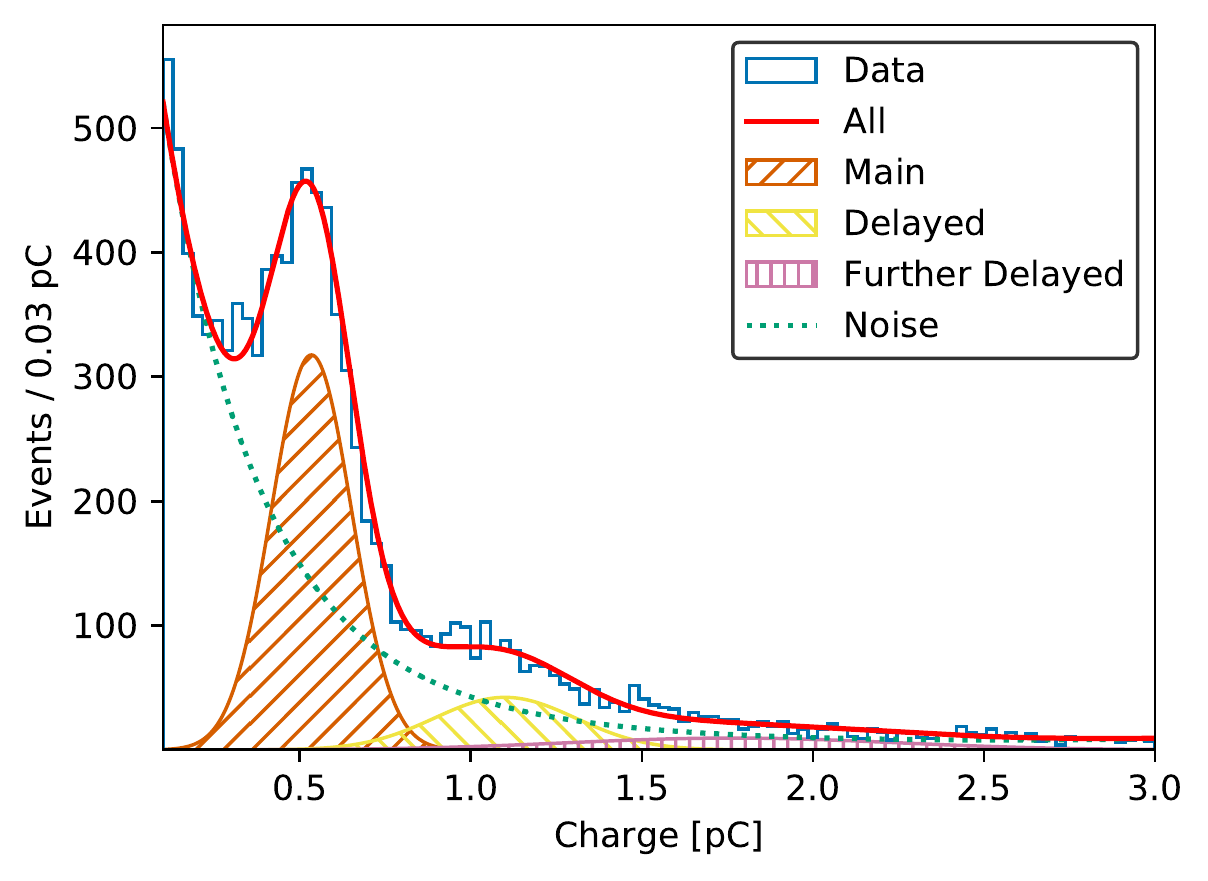}
	\caption{Output charge distribution (blue histogram) with the fitted functions overlaid (three Gaussian functions for the main, delayed and further delayed laser signals, the green dotted line of the exponential function for the random noise, and the red solid line for the combination).
		The net output charge without amplification is shown.}
	\label{fig:charge}
\end{figure}

Figure~\ref{fig:charge} shows the output charge distribution.
The random noise component is determined by fitting an exponential function on the events prior to the laser timing.
Three Gaussian functions together with this fixed exponential one fit well with the output charge distribution for the laser timing events.
The means of the Gaussian functions are $0.534 \pm 0.003$~pC, $1.10 \pm 0.02$~pC, and $1.76 \pm 0.06$~pC.
The first two of them correspond to the main and delayed laser signals.
The gain of the main laser signals is estimated to be $3.3\times 10^6$.
The peaking distribution of the output charge well-separated from the pedestal is different from that of RPCs, which looks closer to the distribution of the random noise.
It could be attributed to the fixed drift gap from the photocathode to the resistive plate and may indicate a photon counting capability.
The doubled output charge for the delayed laser signals bears out the above-mentioned idea of two overlapping pulses due to photon feedback to the photocathode.
The third Gaussian component could indicate three or more overlapping pulses as the mean of the Gaussian is nearly triple, and the measured $t_\mathrm{sig}$ is further delayed.
The ratio of the area under the Gaussian functions is $1 : (0.23 \pm 0.02) : (0.12 \pm 0.01)$, from which the expected rate of photon feedback occurrence under assumption of a Poisson model is estimated to be $0.30 \pm 0.02$.

Given a digitizer with a finer sampling, the overlapping pulses could be resolved and $t_\mathrm{sig}$ could be deduced only from the initial pulse to eliminate the artifactitious delay.
Thus, we evaluate the time resolution of the GasPM prototype only using the main laser signals in the next section.

Signals with much larger output charges due to formation of streamers were also observed.
The streamer pulses exceeding the digitizer input range of $-500$~mV appeared a few tens nanoseconds after the precursor avalanche pulse.
The fraction of signals containing such streamer pulses was about 5\%.
The streamer pulses rarely affect the time resolution since $t_\mathrm{sig}$ is determined from the preceding avalanche pulses.

\subsection{Single-photon time resolution}

\begin{figure*}[tb]
	\begin{minipage}{0.49\textwidth}
		\centering 
		\includegraphics[width=\textwidth]{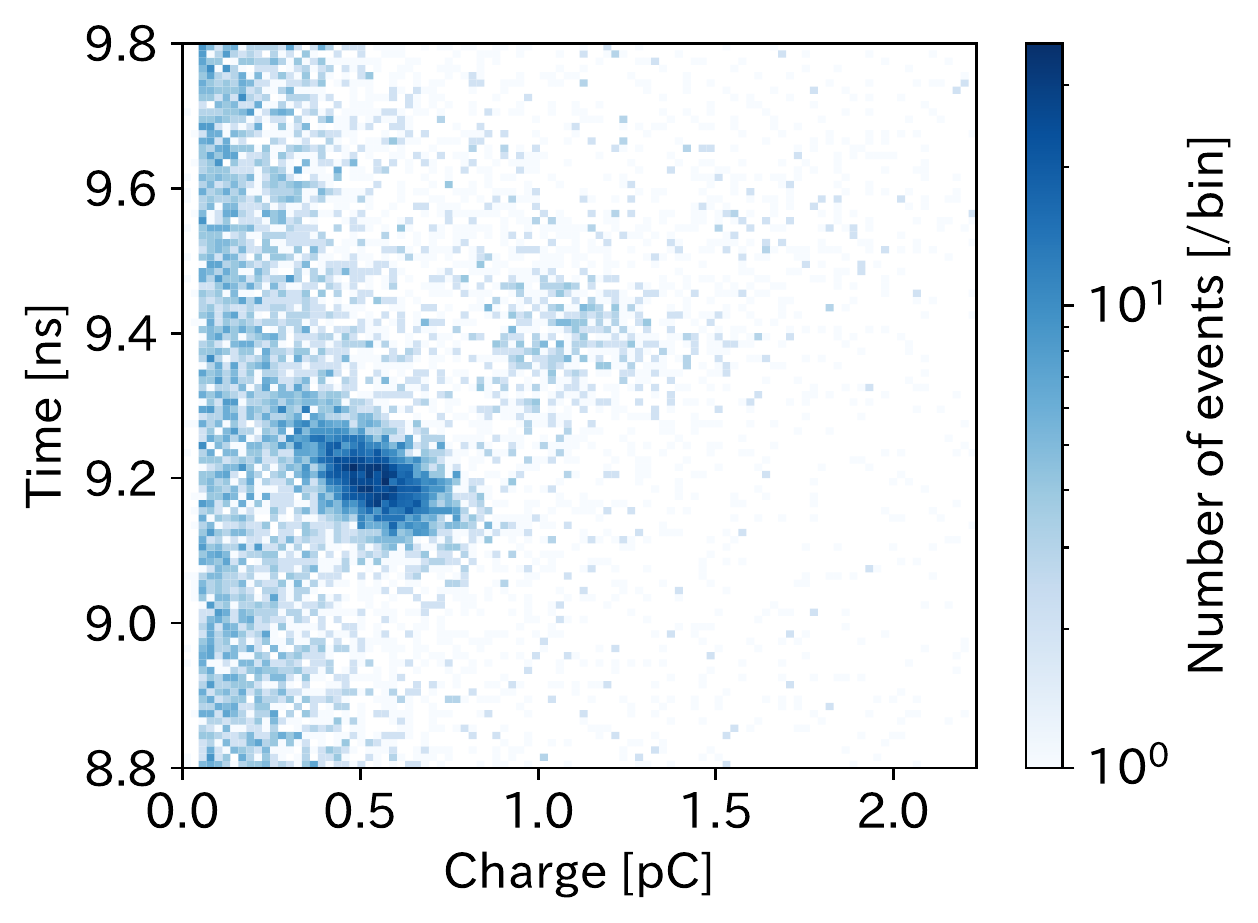}
	\end{minipage}
	\hspace{0.02\textwidth}
	\begin{minipage}{0.49\textwidth}
		\centering 
		\includegraphics[width=\textwidth]{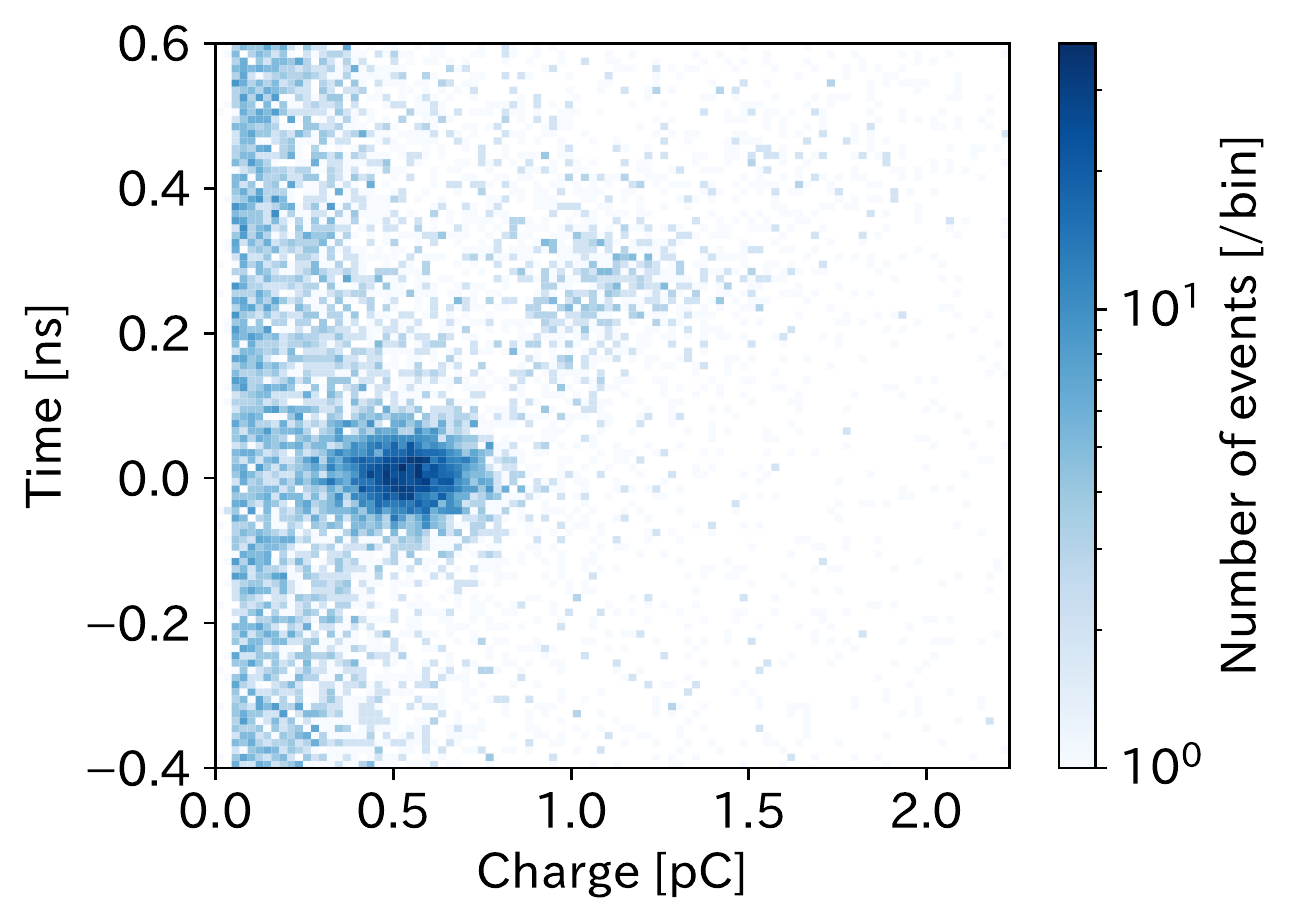}
	\end{minipage}
	\caption{Output charge and time distribution of the GasPM signals without (left) and with (right) the time-charge correction.
		The net output charge without amplification is shown.}
	\label{fig:Q-t}
\end{figure*}

\begin{figure}[tb]
    \centering
    \includegraphics[width=0.5\textwidth]{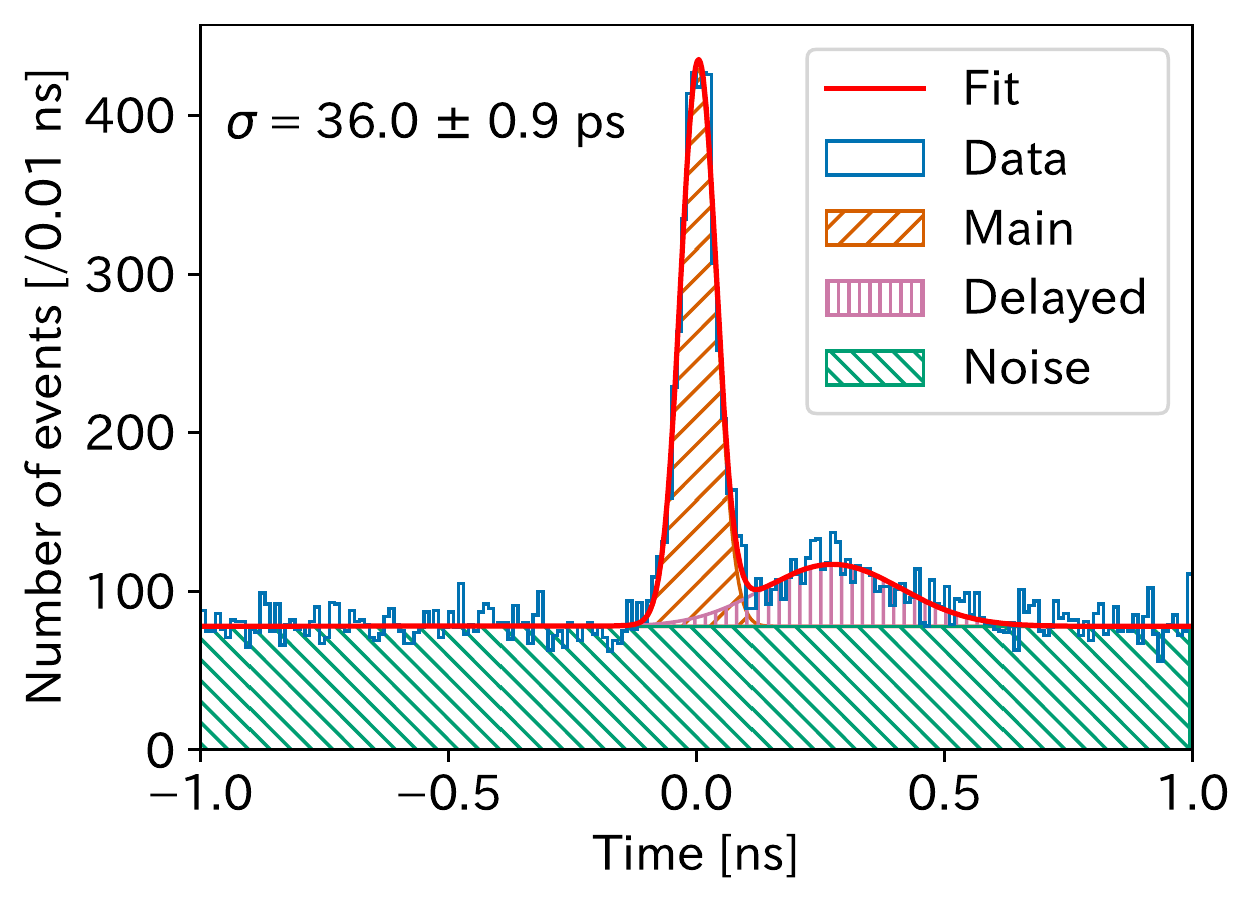}
    \caption{Time distribution of the GasPM signals after the time-charge correction.
    There are two peaks corresponding to the main and delayed laser signals on top of the random noise.
	The fitting result representing each component is superimposed.}
    \label{fig:time_resolution}
\end{figure}

Figure~\ref{fig:Q-t} shows the output charge and time distribution for the GasPM signals.
The main and delayed laser signals cluster around 9.2 and 9.4~ns, respectively. 
Looking at the main laser signals, there is a dependence of the time on the output charge.
It is because the risetime of the signal pulse is not constant but depends on the output charge.
It indicates that the electron drift time could be affected by the space charge.
The dependence is corrected with the following formula,
\begin{equation}
    t_\mathrm{sig} = \frac{a}{\sqrt{Q-b}}+c,
    \label{eq:tQ_correction}
\end{equation}
where $Q$ is the output charge, and $a$, $b$ and $c$ are determined by fitting this formula only on the main laser signals.
The corrected distribution is shown in Fig.~\ref{fig:Q-t} (right), and the projection to the time axis is shown in Fig.~\ref{fig:time_resolution}.
There are two peaks corresponding to the main and delayed laser signals.
The second peak includes the further delayed laser signals.
They are fitted with a double Gaussian plus a constant term, and the time resolution of the main laser signals is measured to be $36.0 \pm 0.9$~ps.
Excluding the laser pulse width and the time resolution of the readout system, the intrinsic time resolution of the GasPM prototype is estimated to be $25.0 \pm 1.1$~ps.

\begin{figure}[tb]
	\centering
	\includegraphics[width=0.5\textwidth]{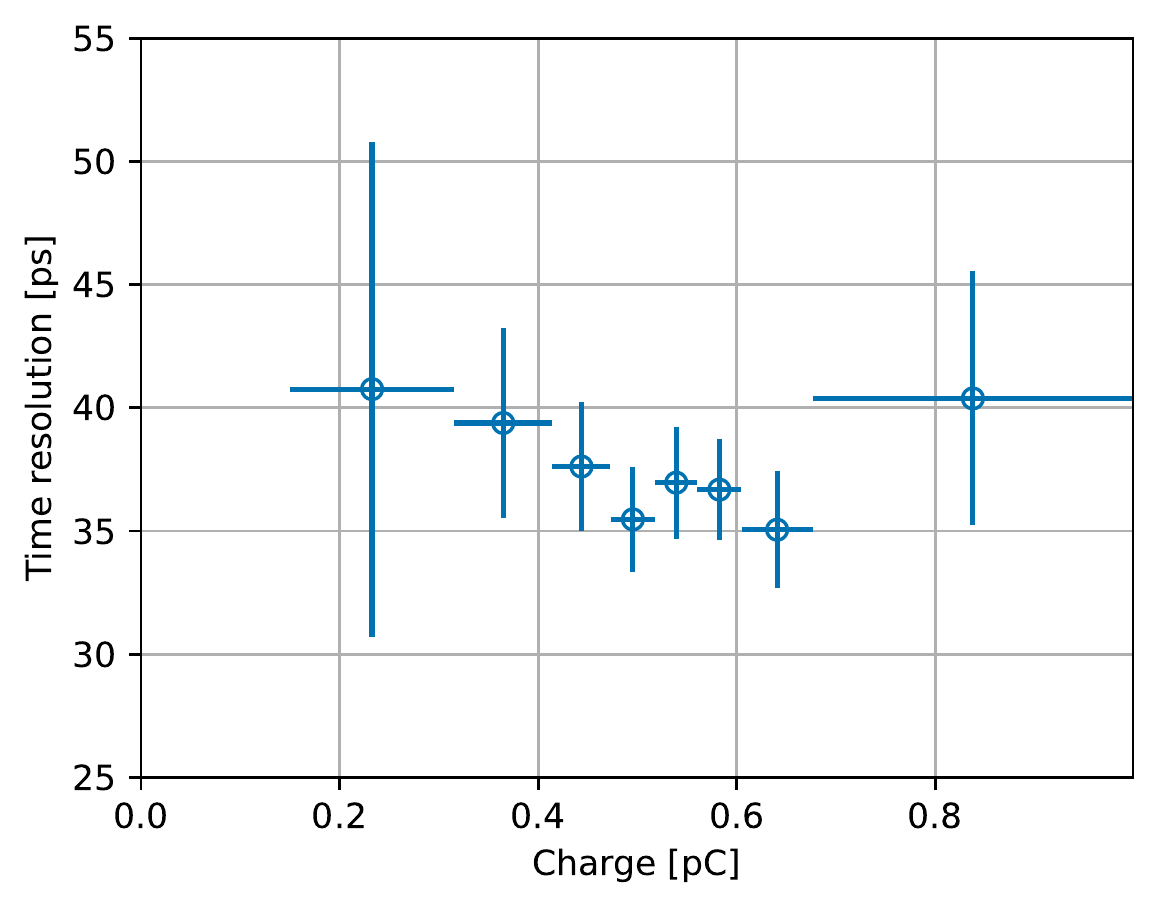}
	\caption{Time resolution of the GasPM prototype at each bin of the output charge.
	The points represent the time resolution including the laser pulse width and the time resolution of the readout system.}
	\label{fig:res_charge}
\end{figure}

Figure~\ref{fig:res_charge} shows the dependence of the time resolution of the GasPM prototype on the output charge.
The best time resolution is $35.0\pm2.4$~ps including laser pulse width and the time resolution of the readout system, or $23.5 \pm 2.5$~ps excluding them.
The time resolution is worse at lower output charges because of a lower signal-to-noise ratio. 


\section{Conclusion}
We shine a light on the RPC-based photosensitive gaseous detector, GasPM, for future nuclear and particle experiments.
It could in principle have a large photocoverage and an excellent single-photon time resolution at low cost, except for other perspective concerns that need to be addressed for practical application.
As a first step, to prove the principle of GasPM, we made the GasPM prototype composed of the LaB$_6$ photocathode and the resistive plate of TEMPAX Float glass.
An electric field of 176~kV/cm was applied to the 170~{\textmu}m gap filled with 90\% R134a and 10\% SF$_6$.
We tested the GasPM prototype with the picosecond pulse laser and demonstrated that it has an intrinsic time resolution of $25.0 \pm 1.1$~ps, which is better than that of MCP-PMTs, at the gain of $3.3\times 10^6$.
Some delayed signals were observed.
The distinct pulse shape of those delayed signals indicates an overlap of successive pulses due to photon feedback from the initial avalanche to the photocathode.
The photon and ion feedback would be more problematic with a photocathode like CsI of a practical quantum efficiency.
It would be a major challenge of GasPM to be addressed in our future development.

\section*{Acknowledgments}
This work was supported by DAIKO FOUNDATION and MEXT/JSPS KAKENHI Grant Numbers JP26610068, JP16H00865, JP19H05099, and JP21H01091.




\begin{thebibliography}{00}

\bibitem{photodetectors}
P.~Kri\v{z}an and S.~Korpar, Photodetectors in Particle Physics Experiments, Annual Review of Nuclear and Particle Science, 63 (2013) 329.


\bibitem{SPAD-QC}
F.~Nolet, et al., Quenching circuit and spad integrated in cmos 65~nm with 7.8~ps fwhm single photon timing resolution, Instruments, 2 (2018) 19.

\bibitem{MCP-PMT_Belle2}
K.~Matsuoka on behalf of the Belle~II TOP group, Performance of the MCP-PMTs of the TOP Counter in the First Beam Operation of the Belle II Experiment, JPS Conference Proceedings, 27 (2019) 011020.
\bibitem{MCP-PMT_Photonis}
D.~A.~Orlov, et al., High collection efficiency MCPs for photon counting detectors, Journal of Instrumentation, 13 (2018) C01047.
\bibitem{MCP-PMT_Photek}
T.~M.~Conneely, et al., The TORCH PMT: a close packing, multi-anode, long life MCP-PMT for Cherenkov applications, Journal of Instrumentation, 10 (2015) C05003.
\bibitem{LAPPD}
A.~V.~Lyashenko, et al., Performance of Large Area Picosecond Photo-Detectors (LAPPD$^\mathrm{TM}$), Nuclear Instruments and Methods in Physics Research Section A, 958 (2020) 162834.

\bibitem{SNSPD}
G.~N.~Gol'tsman, et al., Picosecond superconducting single-photon optical detector, Applied Physics Letters, 79 (2001) 705.
\bibitem{SNSPD_2.6ps}
B.~Korzh, et al., Demonstration of sub-3~ps temporal resolution with a superconducting nanowire single-photon detector, Nature Photonics, 14 (2020) 250.
\bibitem{SNSPD_kilopixel}
E.~E.~Wollman, et al., Kilopixel array of superconducting nanowire single-photon detectors, Optics Express, 27 (2019) 35279.

\bibitem{photosensitive_gas}
R.~Arnold, et al., A fast-cathode pad-photon detector for Cherenkov ring imaging, Nuclear Instruments and Methods in Physics Research Section A, 314 (1992) 465.
\bibitem{PGD}
T.~Francke and V.~Peskov, Photosensitive gaseous detectors and their applications, Nuclear Instruments and Methods in Physics Research Section A, 525 (2004) 1.
\bibitem{PICOSEC-Micromegas}
L.~Sohl, et al., Single photoelectron time resolution studies of the PICOSEC-Micromegas detector, Journal of Instrumentation, 15 (2020) C04053.
\bibitem{PPAC_CsI}
G.~Charpak, et al., Investigation of operation of a parallel-plate avalanche chamber with a CsI photocathode under high gain conditions, Nuclear Instruments and Methods in Physics Research Section A, 307 (1991) 63.
\bibitem{RPC_CsI_1}
P.~Carlson, et al., Beyond the RICH: innovative photosensitive gaseous detectors for new fields of applications, Nuclear Instruments and Methods in Physics Research Section A, 502 (2003) 189.
\bibitem{RPC_CsI_2}
T.~Francke, et al., High rate (up to $10^5$~Hz/cm$^2$), high position resolution (30~{\textmu}m) photosensitive RPCs, Nuclear Instruments and Methods in Physics Research Section A, 533 (2004) 163.
\bibitem{RPC_CsI_3}
P.~Fonte, et al., Novel single photon detectors for UV imaging, Nuclear Instruments and Methods in Physics Research Section A, 553 (2005) 30.
\bibitem{QE_field}
A.~Breskin, CsI UV photocathodes: history and mystery, Nuclear Instruments and Methods in Physics Research Section A, 371 (1996) 116.

\bibitem{1stRPC}
R.~Santonico and R.~Cardarelli, Development of resistive plate counters,  Nuclear Instruments and Methods in Physics Research, 187 (1981) 377.
\bibitem{LaB6}
J.~M.~Lafferty, Boride cathodes, Journal of Applied Physics, 22 (1951) 299.
\bibitem{LaB6_surface}
R.~Nishitani, et al., Surface structures and work functions of the LaB$_6$ (100), (110) and (111) clean surfaces, Surface Science, 93 (1980) 535.
\bibitem{LaB6_oxidized}
B.~Goldstein and D.~J.~Szostak, Characterization of clean and oxidized (100)LaB$_6$, Surface Sicence, 74 (1978) 461.
\bibitem{LaB6_QE}
P.~G.~May, et al., Photoemission from thin-film lanthanum hexaboride, Applied Physics Letters, 57 (1990) 1584.
\bibitem{absorptance}
O.~Deparis, Poynting vector in transfer-matrix formalism for the calculation of light absorption profile in stratified isotropic optical media, Optics Letters, 36 (2011) 3960.
\bibitem{index_LaB6}
H.~Yuan, et al., Size dependent optical properties of LaB$_6$ nanoparticles enhanced by localized surface plasmon resonance, Journal of Rare Earths, 31 (2013) 1096.
\bibitem{index_ITO}
HORIBA technical note TN08: Spectroscopic ellipsometry (2006), available at \url{https://www.horiba.com/fileadmin/uploads/Scientific/Downloads/OpticalSchool_CN/TN/ellipsometer/Lorentz_Dispersion_Model.pdf} (accessed on February 20, 2023).
\bibitem{LaB6_sputtering_power_flow}
C.~M.~Zimmer, et al., Photoemission properties of LaB$_6$ thin films for the use in PIDs, 14th IEEE International Conference on Nanotechnology, 2014, 877.
\bibitem{LaB6_sputtering_heat}
D.~Wang, et al., Effect of heat treatment on the properties of dc magnetron sputtered LaB$_6$/ITO films, Applied Surface Science, 257 (2011) 6418.
\bibitem{RPC_SF6}
P.~Camarri, et al., Streamer suppression with SF$_6$ in RPCs operated in avalanche mode, Nuclear Instruments and Methods in Physics Research Section A, 414 (1998) 317.

\bibitem{RPC_simulation}
W.~Riegler, C.~Lippmann, R.~Veenhof, Detector physics and simulation of resistive plate chambers, Nuclear Instruments and Methods in Physics Research Section A, 500 (2003) 144.
\bibitem{Magboltz2}
S.~F.~Biagi, Monte Carlo simulation of electron drift and diffusion in counting gases under the influence of electric and magnetic fields, Nuclear Instruments and Methods in Physics Research Section A, 421 (1999) 234.
\bibitem{Garfield++}
Garfield++, a toolkit for the detailed simulation of particle detectors based on ionisation measurement in gases and semiconductors, available at \url{https://garfieldpp.web.cern.ch/garfieldpp/} (accessed on February 20, 2023).
\bibitem{RPC_efficiency}
M.~Abbrescia, et al., Progresses in the simulation of Resistive Plate Chambers in avalanche mode, Nuclear Physics B (Proceedings Supplements), 78 (1999) 459.

\bibitem{DRS4}
S.~Ritt, Design and performance of the 6~GHz waveform digitizing chip DRS4, 2008 IEEE Nuclear Science Symposium Conference Record, Dresden, Germany, 2008, 1512.

\end{thebibliography}


\end{document}